# Transverse Magneto-Optical Kerr Effect in Active Magneto-Plasmonic Structures

OLGA BOROVKOVA,[1,*] ANDREY KALISH,[1,2] VLADIMIR BELOTELOV[1,2]

[1]*Russian Quantum Center, Skolkovo 143025, Russia*
[2]*Faculty of Physics, M.V. Lomonosov Moscow State University, Moscow 119991, Russia*
*Corresponding author: o.borovkova@rqc.ru*

**We propose a novel method to enhance the transverse magneto-optical Kerr effect (TMOKE) in the magneto-plasmonic (MP) nanostructures by means of the active dielectric layer. We report the theoretical analysis of the magnetoplasmonic structure with a ferromagnetic dielectric doped with rear-earth ions (Nd³⁺) as the example of a gain layer. The enhancement takes place near the surface plasmon polariton (SPP) resonances of the nanostructures. The stimulated emission of the dopants in the field of SPP wave partially compensates its losses. It is shown that due to a decrease of SPP damping a Q-factor of the MP resonance increases and the TMOKE is increased in comparison with the passive nanostructure.**

Magneto-optical (MO) effects find diverse applications in modern data storage devices and optical isolators [1]. They reveal themselves as a rotation of light polarization due to a medium magnetization or a change of the reflected or transmitted light intensity caused by the medium remagnetization [2].

In this Letter the transverse MO intensity effect is addressed. It takes place when the magnetization is in-plane and is perpendicular to the plane of light incidence, and is usually referred as transverse MO Kerr effect or TMOKE [2]. Thin smooth films of the ferromagnetic metals show noticeable TMOKE in reflection, but they are opaque in transmission. At the same time, the thin smooth films of ferromagnetic dielectrics are transparent, but TMOKE in them is hardly measurable.

Excitation of surface plasmon-polaritons (SPPs) at the surface of a ferromagnetic film might significantly enhance the MO response [3-6]. Intensified MO effects were demonstrated in noble-/ ferromagnetic-metal heterostructures [7-9], and in perforated ferromagnetic-metal films [10-13].

In Refs. [14-16] the enhanced MO effect in magneto-plasmonic (MP) crystals composed of a ferromagnetic dielectric covered by a noble metal grating was reported. This structure combines a strong magnetic response with a high light transmission at the wavelengths greater than 650 nm. Thus, it is possible to observe the intensity MO effect in transmitted light as well as in the reflected one. It should be stressed, that in the MP crystal the relative change in the intensity of transmitted light can be two orders of magnitude higher than for smooth ferromagnetic metal films [14]. The value of TMOKE depends on the Q-factor of MO resonance, that, in its turn, is limited by structure's losses.

Although, ferromagnetic dielectrics possess smaller absorption than ferromagnetic metals, they still have noticeable losses due to the presence of iron ions [17]. Besides that, optical power dissipates in the metal parts of the MP structure due to inelastic processes associated with the motion of conduction electrons in metals. In the MP nanostructures the SPPs suffer from the strong dissipation in both noble metal film and ferromagnetic dielectric and, thus, a Q-factor of the MO resonance decreases.

Loss compensation and amplification of SPPs have been investigated in gain plasmonics for active non-magnetic materials [18-27]. The SPP propagates along the interface of metal and dielectric material doped by active centers. The pump beam illuminates dielectric, and excites the active centers in the gain medium. An electromagnetic field of the SPP stimulates an emission of activated ions or molecules. This radiation has the same phase and polarization as the surface wave, and effectively amplifies the SPP. The SPP amplification have been demonstrated in the solutions of dyes (cresyl violet and rhodamine 101 [21]), PMMA doped with dyes [22] or quantum dots [23]. Lasing action was investigated in different types of plasmonic crystals incorporating optically-pumped gain media, like arrays of metal holes [24], strongly coupled plasmonic nanocavity arrays [25], and in periodic arrays of metallic nanowires embedded in the gain medium [26]. However, all these materials are non-magnetic and aren't suitable for an enhancement of the magneto-optical effects.

Gain materials are successfully applied for spasers, the subdiffractive sources of linearly polarized light [27, 28]. Recently the MO spaser has been suggested [29]. It combines amplifying core with magnetic shell and provides the coherent circularly polarized light unachievable in nonmagnetic spasers.

In this paper, we propose a novel way to enhance MO effects due to the layer of gain medium. It can be realized, for example, in the MP crystal with the ferromagnetic dielectric material doped by active centers. Here, an enhancement of TMOKE is addressed theoretically for different values of the gain parameter and for the various thicknesses of the gain media in the MP crystal. Also it is shown that partial loss compensation of SPP leads to an increase of magneto-plasmonic resonance Q-factor.

Bismuth-substituted yttrium iron garnet (BIG) possesses strong MO response [2]. We choose for our simulations the rear-earth ions Nd³⁺ as active centers. Magnetic properties of Nd-doped YIG have been studied in Ref. [30], however, Nd ions were not employed as active centers. The MO response of BIG is rather strong at the Nd emission wavelength, and this dopant is well-known due to its wide application in solid state lasers, where activated non-magnetic garnets are used. The BIG film doped with the rear-earth ions can be grown on the suitable substrate by the standard isothermal liquid-phase epitaxy method. Another possible way is to create the doped ferromagnetic film by the RF magnetron sputtering method [31].

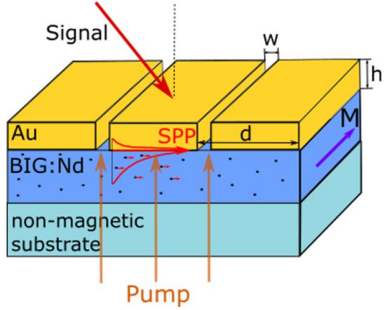

Fig. 1. The scheme of the MP crystal. $d$ = 900 nm, $h$ = 150 nm, and $w$ = 100 nm.

We consider the MP crystal of the 1D gold grating deposited on the doped BIG grown on top of the non-magnetic gadolinium gallium garnet (GGG) substrate (see Fig. 1). Here $d$ is a period of the gold grating, $h$ refers to its depth, and $w$ denotes the air groove width. We employ Voigt geometry, when the incidence plane of the $p$-polarized light is perpendicular to the magnetization **M**. The following parameters of the MP structure are chosen: $d$ = 900 nm, $h$ = 150 nm, and $w$ = 100 nm. The depth of active layer is $2.5$ μm. The gold layer is characterized by a dielectric function $\varepsilon_1$. The gain in the film is determined by the imaginary part of dielectric function $\varepsilon_2 = \varepsilon_2' + i\varepsilon_2''$, where $\varepsilon_2'' < 0$. An emission of dopants has Lorentz line shape. Optical pumping of neodymium can be operated at $0.8$ μm, and it has one of the luminescence peaks at $1.06$ μm [32]. MO properties of BIG are described by a gyration vector **g** = $a$**M** [2], at the dopants emission wavelength the gyration is about $6 \cdot 10^{-4}$.

A pump beam illuminates the structure from below, comes through the substrate, and excites dopants inside the ferromagnetic film. The absorption of pump beam in the thin substrate film is negligibly small. At the same time the signal beam is obliquely incident from above to excite the SPP wave along the metal/dielectric interface. Varying incidence angle of the signal beam one can adjust the frequency of the SPP excitation to the dopant emission frequency. The stimulated emission from the active centers diminishes optical losses and thus increases the SPP propagation length and the Q-factor of plasmonic resonance.

Let us consider this problem in details. The SPP is excited if its wavevector, $k_{SPP}$, is matched with the in-plane component, $k_\parallel$, of the incident light wavevector through the momentum acquired from the gold grating, $2\pi m/d$: $k_\parallel = k_{SPP} + m\frac{2\pi}{d}$, where $m$ is an integer.

Within the empty-lattice approximation $k_{SPP}$ can be estimated by the one for the smooth metal/magnetic dielectric interface [14]:

$$k_{SPP} \approx \frac{\omega}{c}\sqrt{\frac{\varepsilon_1\varepsilon_2}{\varepsilon_1+\varepsilon_2}}(1+\alpha g), \quad (1)$$

where $\omega$ is the incident light frequency, $c$ is the light velocity in vacuum, and parameter $\alpha = \left(-\varepsilon_1\varepsilon_2\right)^{-\frac{1}{2}}\left(1-\varepsilon_2^2/\varepsilon_1^2\right)^{-1}$.

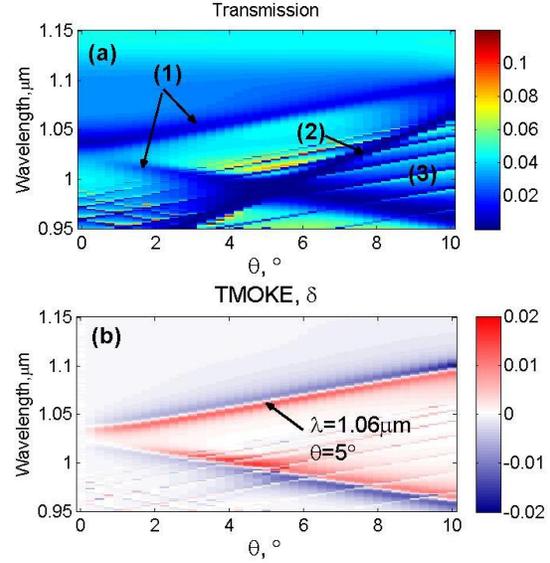

Fig. 2. Transmission spectrum (a) and the TMOKE (b) of the MP structure as a function of light wavelength and incident angle.

SPP excitation in the structure manifests itself as resonances in transmission spectrum. On the basis of the rigorous coupled waves analysis (RCWA) [33,34] spectral transmittance of the structure was found (Fig. 2(a)). Its peculiarities point out to the different optical surface modes. We are mostly interested in the Fano resonances related to the SPP at the gold/iron-garnet interface (resonances are denoted with (1)). Besides that, in the figure there are the Fano resonance referred to the SPP at the air/gold interface (denoted with (2)) and a set of waveguide modes (denoted with (3)).

The influence of the magnetization on the optical properties of the sample in TMOKE can be characterized by the relative change in the transmitted light intensity T(**M**) when the structure is remagnetized

$$\delta = 2\frac{T(\mathbf{M})-T(-\mathbf{M})}{T(\mathbf{M})+T(-\mathbf{M})}. \quad (2)$$

The most pronounced resonances of $\delta$ emerge with the excitation of SPP waves at the gold/iron-garnet interface (Fig. 2(b)), and, naturally, they depend on the dielectric properties of the BIG film.

We've fitted the transmission resonances by Fano resonance shape [35] for various incident angles and found out that at $\theta = 5°$ the central wavelength is $1.06$ μm. Thus, SPP waves excited at these incident angle could be enhanced by emission of Nd ions.

The gain properties of Nd-doped BIG films haven't been studied yet, Thus, for our simulations we have to take the known values of the emission and absorption properties of the Nd ions in the other hosting material. We choose the yttrium aluminium garnet (YAG) that is justified as follows. YAG has crystal lattice parameter close to BIG. The emission wavelengths for Nd-doped BIG and YAG might be slightly different (by about 10 nm [32]). However, such difference can be easily compensated by the variation of the incident angle. Moreover, this wavelength shift might change the absorption in YIG by only about 5 cm$^{-1}$ [2]. The exact values of absorption and emission parameters of the Nd-doped BIG could be determined in the direct measurements of the sample. In the passive BIG film, the absorption is about 30 cm$^{-1}$ at 1.06 μm, that corresponds to $\varepsilon_2'' \approx 0.001$. For our simulations we assume that Nd concentration in BIG film is $6 \cdot 10^{20}$ cm$^{-3}$, the estimated required pump power is about 15 mW. The decay rate of the excited Nd$^{3+}$ ions is 0.23 ms. We took emission cross section about $5 \cdot 10^{-19}$ cm$^2$ that was reported for Nd ions in [32] and refs. inside. This yields gain 300 cm$^{-1}$ that corresponds $\varepsilon_2'' \approx -0.01$ at the emission wavelengths. So, we consider the impact of gain in the magnetic dielectric up to this value.

In Fig. 3 one can see spectra of the transmittance and TMOKE for various values of gain, i.e. $\varepsilon_2''$. On each panel there are two groups of curves. Left groups refer to the resonant case, when the MO effect occurs at the emission wavelength of dopants (incident angle $\theta = 5°$). Passive structure, when there is no optical pump and $\varepsilon_2''$ is positive, is indicated with the blue curves. One can see that the resonance in transmission gets sharper with the increase of the gain parameter (Fig. 3a). The MO parameter δ behaves similarly, the resonance becomes sharper with the gain and its maximum value increases (Fig. 3b). On the contrary, when the incident angle is taken to provide MO effect at the wavelength, shifted from the dopants' emission wavelength (right groups of curves) we do not observe any enhancement either for transmission or for MO parameter. However, one can see a slight increase of transmission at 1.06 μm due to the SPP field amplification. Thus, we observe resonance enhancement of MO effect due to the adjustment of the emission wavelength and MO resonance. One can see that the maximum value of δ rises by almost 1.5 times for $\varepsilon_2'' = -0.01$ in comparison with the passive plasmonic structure (see Fig. 3b).

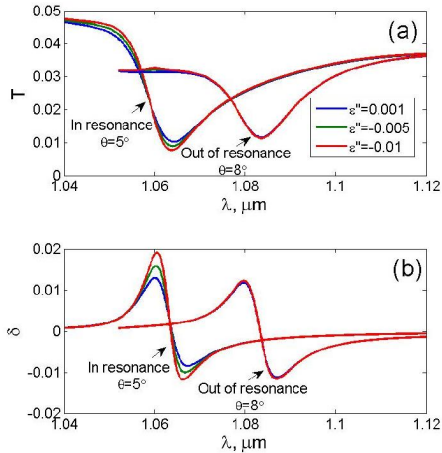

Fig. 3. (a). Transmission spectrum of the plasmonic structure versus the wavelength at the various gain levels. (b). Dependence of the TMOKE parameter δ on wavelength for three gain levels.

The observed enhancement of TMOKE in active MP structure is explained as follows. SPP losses are compensated by gain in the iron-garnet film. A quality factor of resonance is inversely proportional to the damping of a system, so the reduction of losses leads to an increase of a quality factor of the transmission resonance. Moreover, by its definition, $Q = \omega_0 / \Delta\omega$, where $\omega_0$ is a central frequency and $\Delta\omega$ is a width of transmission resonance. Thus, a resonance with higher quality is steeper and has larger derivative $\partial T / \partial \lambda$. On the other hand, it was shown in [15] that the TMOKE enhancement can be explained by the magnetization induced shift of the SPP resonance frequency and can be represented as

$$\delta \sim \frac{1}{T} \cdot \frac{\partial T}{\partial \lambda} \qquad (3)$$

Thus, TMOKE parameter δ grows with an increase of gain.

For incidence angle $\theta = 5°$ we have calculated Q-factor of transmission resonance versus gain parameter and have compared them with the curve $\delta(\varepsilon_2'')$. It depends almost linearly on the gain parameter $\varepsilon_2''$ (see Fig. 4a). The Q-factor of transmittance resonance reveals an increase by 13% at the gain parameter $\varepsilon_2'' = -0.01$. However, the MO effect grows by 1.5 times in comparison with the passive MP crystal. Therefore, we can conclude that in this structure the MO effect is very sensitive to the imaginary part of dielectric permittivity.

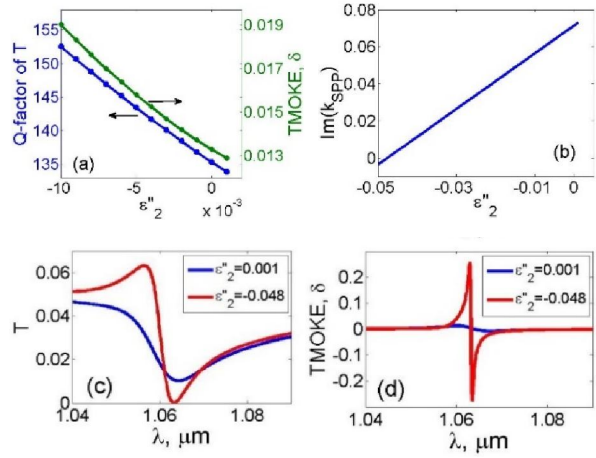

Fig. 4. (a) Dependence of the Q-factor of the transmission resonance and TMOKE on the gain for the MP crystal. (b) Imaginary part of $k_{SPP}$ versus the gain parameter. (c) Transmission spectrum and (d) TMOKE in passive MP structure (blue) and in case of full loss compensation (red).

The gain affects the propagation constant of the SPP, mostly its imaginary part, $\mathrm{Im}(k_{SPP})$ (see Eq. (1)). Expectedly, $\mathrm{Im}(k_{SPP})$ decreases as the gain arises (Fig. 4(b)). One can see that $\varepsilon_2'' = -0.01$ gives just partial compensation of the SPP losses. Full compensation requires the gain coefficient almost 5 times stronger, $\varepsilon_2'' = -0.048$. In this case we can see that the transmission at resonance dips to almost zero, and both resonances of T and δ have very steep profile (Fig. 4(c-d)). TMOKE becomes 10 times larger than for $\varepsilon_2'' = -0.01$.

It is well-known, that the SPPs are strongly localized at the metal/dielectric interface and their intensity exponentially decays away from the border. Thus, the stimulated emission of the rear-earth ions in the external field of the SPP wave takes place only near the metal/dielectric interface and strongly depends on the gain layer thickness. We address an increase of the TMOKE versus the thickness of the gain layer for two fixed values of the gain parameter, $\varepsilon_2'' = -0.01$ and $\varepsilon_2'' = -0.005$ (Fig. 5). The steepest increase of the effect takes place inside the ~200nm-thick gain layer where strong localization of the electromagnetic field occurs. Moving away from the metal/dielectric interface the SPP field decreases exponentially (see the inset in Fig. 5 with the field distribution inside the magneto-plasmonic structure) and the stimulated emission decays. For the 600 nm-thick gain layer the TMOKE gets to the plateau and further it varies only negligibly. It is almost equal to a penetration depth of the SPP that can be estimated as $h_{SPP} = 1/|k_z| = 585$ nm for the Au/BIG interface (vertical dashed line in Fig. 5). Therefore, the effective enhancement of TMOKE occurs just in the 600nm-thick layer near the metal/dielectric interface, the layer where the SPP exists. The dopants located at a larger depth of the ferromagnetic dielectric do not impact on the TMOKE.

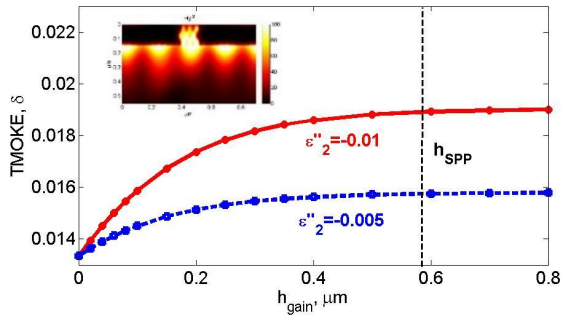

Fig. 5. TMOKE versus thickness of the active ferromagnetic medium. Solid lines show the values of δ for $\varepsilon''_2 = -0.01$, and dashed lines correspond to δ for $\varepsilon''_2 = -0.005$. Vertical dashed line ($h_{SPP}$) shows a penetration depth of SPP wave in the BIG film. Inset: a field distribution inside the MP structure.

In conclusion, here we report an amplification of the TMOKE in active MP structure by means of the stimulated emission of optically-pumped rear-earth ions. Stimulated emission partially compensates SPP losses that increases the quality of transmission resonances, which, in its turn, is proportional to the TMOKE. For the considered magneto-plasmonic structure TMOKE shows 1.5-time growth for the gain parameter $\varepsilon''_2 \approx -0.01$. Moreover, further compensation of losses provides even steeper resonances and even higher values of the TMOKE. The most impact to the enhancement of TMOKE is provided by dopants in thin layer near metal/dielectric interface, where the electromagnetic field of the SPP penetrates. The reported effect provides a novel method of tunable enhancement of the MO effect in the MP structures. Varying the optical pumping one can change the value of TMOKE.

**Funding.** Russian Foundation for Basic Research (RFBR) (16-32-60135 mol_a_dk, 14-02-01012)

## References

1. J.P. Castera, and T. Suzuki, in *Magneto-optical Devices. The Optics Encyclopedia* (Wiley, 2007).
2. A. Zvezdin, and V. Kotov, in *Modern Magnetooptics and Magnetooptical Materials* (IOP, 1997).
3. P.E. Ferguson, O.M. Stafsudd, and R.F. Wallis, Physica B&C **86-88**, 1403 (1977).
4. R.K. Hickernell, and D. Sarid, Opt. Lett. **12**, 570 (1987).
5. R.D. Olney, R.J. Romagnoli, and P.E. Ferguson, J. Opt. Soc. Am. B **3**, 1526 (1986).
6. N. Bonod, R. Reinisch, E. Popov, and M. Neviere, J. Opt. Soc. Am. B **21**, 791 (2004).
7. J.B. Gonzalez-Diaz, A. Garcia-Martin, G. Armelles, J.M. Garcia-Martin, C. Clavero, A. Cebollada, R.A. Lukaszew, J.R. Skuza, D.P. Kumah, and R. Clarke Phys. Rev. B **76**, 153402 (2007).
8. G. Armelles, A. Cebollada, A. Garcia-Martin, J.M. Garcia-Martin, M.U. Gonzalez, J.B. Gonzalez-Diaz, E. Ferreiro-Vila, and J.F. Torrado, J. Opt. A **11**, 114023 (2009).
9. V.V. Temnov, Nature Photon. **4**, 107 (2010).
10. D.M. Newman, L.M. Wears, R.J. Matelon, and I.R. Hooper, J. Phys.: Condens. Matter **20**, 345230 (2008).
11. E.Y. Buchin, E.I. Vaganova, V.V. Naumov, V.A. Paporkov, and A.V. Prokaznikov, Tech. Phys. Lett. **35**, 589 (2008).
12. C. Clavero, K. Yang, J.R. Skuza, and R.A. Lukaszew, Opt. Lett. **35**, 1557 (2010).
13. N.E. Khokhlov, A.R. Prokopov, A.N. Shaposhnikov, V.N. Berzhansky, M.A. Kozhaev, S.N. Andreev, A.P. Ravishankar, V.G. Achanta, D.A. Bykov, A.K. Zvezdin, and V.I. Belotelov, J. Phys. D: Appl. Phys. **48**, 095001 (2015).
14. V.I. Belotelov, I.A. Akimov, M.Pohl, V.A. Kotov, S. Kasture, A.S. Vengurlekar, A.V. Gopal, D.R. Yakovlev, A.K. Zvezdin, and M. Bayer, Nature Nanotech. **6**, 370 (2011).
15. M. Pohl, L.E. Kreilkamp, V.I. Belotelov, I.A. Akimov, A.N. Kalish, N.E. Khokhlov, V.J. Yallapragada, A.V. Gopal, M. Nur-E-Alam, and M. Vasiliev, New Journal of Physics **15**, 075024 (2013).
16. V.I. Belotelov, L.E. Kreilkamp, A.N. Kalish, I.A. Akimov, D.A. Bykov, S. Kasture, V.J. Yallapragada, A.V. Gopal, A.M. Grishin, S.I. Khartsev, M. Nur-E-Alam, M. Vasiliev, L.L. Doskolovich, D.R. Yakovlev, K.Alameh, A.K. Zvezdin, and M. Bayer, Phys. Rev. B **89**, 045118 (2014).
17. A.R. Prokopov, P.M. Vetoshko, A.G. Shumilov, A.N. Shaposhnikov, A.N. Kuz'michev, N.N. Koshlyakova, V.N. Berzhansky, A.K. Zvezdin, V.I. Belotelov, Journal of Alloys and Compounds **671**, 403 (2016).
18. P. Berini, and I. De Leon, Nature Photon. **6**, 16 (2012).
19. G. Plotz, H. Simon, and J. Tucciarone, J. Opt. Soc. Am. **69**, 419 (1979).
20. A.N. Sudarkin, and P.A. Demkovich, Sov. Phys. Tech. Phys. **34**, 764 (1988).
21. J. Seidel, S. Grafstorm, and L. Eng, Phys. Rev. Lett. **94**, 177401 (2005).
22. M.A. Noginov, V.A. Podolshkiy, G. Zhu, M. Mayy, M. Bahoura, J.A. Adegoke, B.A. Ritzo, and K. Reynolds, Opt. Express **16**, 1385 (2008).
23. P.M Bolger, W. Dickson, A.V. Krasavin, L. Liebscher, S.G. Hickey, D.V. Skryabin, and A.V. Zayats, Opt. Lett. **35**, 1197 (2010).
24. F. van Beijnum, P.J. van Veldhoven, E.J. Geluk, M.J.A. de Dood, G. W. Hooft, and M.P. van Exter, Phys. Rev. Lett. **110**, 206802 (2013).
25. W. Zhou, M. Dridi, J.Y. Suh, Ch. H. Kim, D. T. Co, M.R. Wasielewski, G.C. Schatz, and T.W. Odom, Nature Nanotech. **8**, 506 (2013).
26. J. Cuerda, F. Rüting, F.J. Garcia-Vidal, and J. Bravo-Abad, Phys. Rev. B **91**, 041118(R) (2015).
27. D.J. Bergman, and M.I. Stockman, Phys. Rev. Lett. **90**, 027402 (2003).
28. M.A. Noginov, G. Zhu, A.M. Belgrave, R. Bakker, and V.M. Shalaev, Nature **460**, 1110 (2009).
29. D.G. Baranov, A.P. Vinogradov, A.A. Lisyansky, Y.M. Strelniker, and D.J. Bergman, Opt. Lett. **38**, 2002 (2013).
30. B.H. Clarke, Phys. Rev **139**, A1944 (1965).
31. E. Devaux, T.W. Ebbesen, J.-C. Weeber, and A. Dereux, Appl. Phys. Lett. **83**, 4936 (2003).
32. M. Birnbaum, and C.F. Klein, J. Appl. Phys. **44**, 2928 (1973).
33. M.G. Moharam, E.B. Grann, D.A. Pommet, T.K. Gaylord, J. Opt. Soc. Am. A **12**, 1068 (1995).
34. L. Li, J. Opt. A Pure Appl. Opt. **5**, 345 (2003).
35. B. Luk'yanchuk, N.I. Zheludev, S.A. Maier, N.J. Halas, P. Nordlander, H. Giessen, and Ch. T. Chong, Nature Mat. **9**, 707 (2010).

**Full References**

1. J.P. Castera, and T. Suzuki, in *Magneto-optical Devices. The Optics Encyclopedia* (Wiley, 2007).
2. A. Zvezdin, and V. Kotov, in *Modern Magnetooptics and Magnetooptical Materials* (IOP, 1997).
3. P.E. Ferguson, O.M. Stafsudd, and R.F. Wallis, Surface magnetoplasma waves in nickel, Physica B&C **86-88**, 1403-1405 (1977).
4. R.K. Hickernell, and D. Sarid, Long-range surface magnetoplasmons in thin nickel films, Opt. Lett. **12**, 570-572 (1987).
5. R.D. Olney, R.J. Romagnoli, and P.E. Ferguson, Optical and magneto-optical effects of surface plasma waves with damping in iron thin films, J. Opt. Soc. Am. B **3**, 1526-1528 (1986).
6. N. Bonod, R. Reinisch, E. Popov, and M. Neviere, Optimization of surface-plasmon-enhanced magneto-optical effects, J. Opt. Soc. Am. B **21**, 791-797 (2004).
7. J.B. Gonzalez-Diaz, A. Garcia-Martin, G. Armelles, J.M. Garcia-Martin, C. Clavero, A. Cebollada, R.A. Lukaszew, J.R. Skuza, D.P. Kumah, and R. Clarke, Surface-magnetoplasmon nonreciprocity effects in noble-metal/ferromagnetic heterostructures, Phys. Rev. B **76**, 153402 (2007).
8. G. Armelles, A. Cebollada, A. Garcia-Martin, J.M. Garcia-Martin, M.U. Gonzalez, J.B. Gonzalez-Diaz, E. Ferreiro-Vila, and J.F. Torrado, Magnetoplasmonic nanostructures: systems supporting both plasmonic and magnetic properties, J. Opt. A **11**, 114023 (2009).
9. V.V. Temnov, Active magnetoplasmonics in hybrid metal/ferromagnet/metal microinterferometers, Nature Photon. **4**, 107-111 (2010).
10. D.M. Newman, L.M. Wears, R.J. Matelon, and I.R. Hooper, Magneto-optic behavior in the presence of surface plasmons, J. Phys.: Condens. Matter **20**, 345230 (2008).
11. E.Y. Buchin, E.I. Vaganova, V.V. Naumov, V.A. Paporkov, and A.V. Prokaznikov, Enhancement of the transversal magnetooptical Kerr effect in nanoperforated cobalt films, Tech. Phys. Lett. **35**, 589-593 (2008).
12. C. Clavero, K. Yang, J.R. Skuza, and R.A. Lukaszew, Magnetic-field modulation of surface plasmon polaritons on gratings, Opt. Lett. **35**, 1557-1559 (2010).
13. N.E. Khokhlov, A.R. Prokopov, A.N. Shaposhnikov, V.N. Berzhansky, M.A. Kozhaev, S.N. Andreev, A.P. Ravishankar, V.G. Achanta, D.A. Bykov, A.K. Zvezdin, and V.I. Belotelov, Photonic crystals with plasmonic patterns: novel type of the heterostructures for enhanced magneto-optical activity, J. Phys. D: Appl. Phys. **48**, 095001 (2015).
14. V.I. Belotelov, I.A. Akimov, M.Pohl, V.A. Kotov, S. Kasture, A.S. Vengurlekar, A.V. Gopal, D.R. Yakovlev, A.K. Zvezdin, and M. Bayer, Enhanced magneto-optical effects in magnetoplasmonic crystals, Nature Nanotech. **6**, 370-376 (2011).
15. M. Pohl, L.E. Kreilkamp, V.I. Belotelov, I.A. Akimov, A.N. Kalish, N.E. Khokhlov, N.J. Yallapragada, A.V. Gopal, m. Nur-E-Alam, and M. Vasiliev, Tuning of the transverse magneto-optical Kerr effect in magneto-plasmonic crystals, New Journal of Physics **15**, 075024 (2013).
16. V.I. Belotelov, L.E. Kreilkamp, A.N. Kalish, I.A. Akimov, D.A. Bykov, S. Kasture, V.J. Yallapragada, A.V. Gopal, A.M. Grishin, S.I. Khartsev, M. Nur-E-Alam, M. Vasiliev, L.L. Doskolovich, D.R. Yakovlev, K.Alameh, A.K. Zvezdin, and M. Bayer, Magnetophotonic intensity effects in hybrid metal-dielectric structures, Phys. Rev. B **89**, 045118 (2014).
17. A.R. Prokopov, P.M. Vetoshko, A.G. Shumilov, A.N. Shaposhnikov, A.N. Kuz'michev, N.N. Koshlyakova, V.N. Berzhansky, A.K. Zvezdin, V.I. Belotelov, Epitaxial BieGdeSc iron-garnet films for magnetophotonic applications, Journal of Alloys and Compounds **671**, 403-407 (2016).
18. P. Berini, and I. De Leon, Surface plasmon-polariton amplifiers and lasers, Nature Photon. **6**, 16-24 (2012).
19. G. Plotz, H. Simon, and J. Tucciarone, Enhanced total reflection with surface plasmons, J. Opt. Soc. Am. **69**, 419-422 (1979).
20. A.N. Sudarkin, and P.A. Demkovich, Excitation of surface electromagnetic waves on the boundary of a metal with an amplifying medium, Sov. Phys. Tech. Phys. **34**, 764-766 (1988).
21. J. Seidel, S. Grafstorm, and L. Eng, Stimulated emission of surface plasmons at the interface between a silver film and an optically pumped dye solution, Phys. Rev. Lett. **94**, 177401 (2005).
22. M.A. Noginov, V.A. Podolshkiy, G. Zhu, M. Mayy, M. Bahoura, J.A. Adegoke, B.A. Ritzo, and K. Reynolds, Compensation of loss in propagating surface plasmon polariton by gain in adjacent dielectric medium, Opt. Express **16**, 1385-1392 (2008).
23. P.M Bolger, W. Dickson, A.V. Krasavin, L. Liebscher, S.G. Hickey, D.V. Skryabin, and A.V. Zayats, Amplified spontaneous emission plasmon polaritons and limitations on the increase of their propagation length, Opt. Lett. **35**, 1197-1199 (2010).
24. F. van Beijnum, P.J. van Veldhoven, E.J. Geluk, M.J.A. de Dood, G. W. Hooft, and M.P. van Exter, Surface Plasmon lasing Observed in Metal Hole Arrays, Phys. Rev. Lett. **110**, 206802 (2013).
25. W. Zhou, M. Dridi, J.Y. Suh, Ch. H. Kim, D. T. Co, M.R. Wasielewski, G.C. Schatz, and T.W. Odom, Lasing action in strongly coupled plasmonic nanocavity arrays, Nature Nanotech. **8**, 506-511 (2013).
26. J. Cuerda, F. Rüting, F.J. Garcia-Vidal, and J. Bravo-Abad, Theory of lasing action in plasmonic crystals, Phys. Rev. B **91**, 041118(R) (2015).
27. D.J. Bergman, and M.I. Stockman, Surface Plasmon Amplification by Stimulated Emission of Radiation: Quantum Generation of Coherent Surface Plasmons in Nanosystems, Phys. Rev. Lett. **90**, 027402 (2003).
28. M.A. Noginov, G. Zhu, A.M. Belgrave, R. Bakker, and V.M. Shalaev, Demonstration a spaser-based nanolaser, Nature **460**, 1110-1112 (2009).
29. D.G. Baranov, A.P. Vinogradov, A.A. Lisyansky, Y.M. Strelniker, and D.J. Bergman, Magneto-optical spaser, Opt. Lett. **38**, 2002-2004 (2013).
30. B.H. Clarke, Rear-Earth Ion Relaxation Time and G Tensor in Rear-Earth-Doped Yttrium Iron Garnet. II. Neodymium, Phys. Rev **139**, A1944 (1965).
31. E. Devaux, T.W. Ebbesen, J.-C. Weeber, and A. Dereux, Launching and decoupling surface plasmons via micro-gratings, Appl. Phys. Lett. **83**, 4936-4938 (2003).
32. M. Birnbaum, and C.F. Klein, Stimulated emission cross section at 1.061um in Nd:YAG, J. Appl. Phys. **44**, 2928 (1973).
33. M.G. Moharam, E.B. Grann, D.A. Pommet, T.K. Gaylord, Formulation for stable and efficient implementation of the rigorous coupled-wave analysis of binary gratings, J. Opt. Soc. Am. A **12**, 1068 (1995).
34. L. Li, Fourier modal method for crossed anisotropic gratings with arbitrary permittivity and permeability tensors, J. Opt. A Pure Appl. Opt. **5**, 345 (2003).
35. B. Luk'yanchuk, N.I. Zheludev, S.A. Maier, N.J. Halas, P. Nordlander, H. Giessen, and Ch. T. Chong, The Fano resonance in plasmonic nanostructures and metamaterials, Nature Mat. **9**, 707-715 (2010).